\documentclass[conference]{IEEEtran}
\IEEEoverridecommandlockouts
\usepackage{cite}
\usepackage{amsmath,amssymb,amsfonts}
\usepackage{algorithmic}
\usepackage{graphicx}
\usepackage{textcomp}
\usepackage{xcolor}
\usepackage{booktabs}
\usepackage{multirow}
\usepackage{csquotes}
\usepackage[inline]{enumitem}
\usepackage{url}
\usepackage{tikz}
\usetikzlibrary{arrows.meta, positioning, shapes.geometric, fit, calc}
\usepackage{pgfplots}
\pgfplotsset{compat=1.18}

\usepackage[
  colorlinks=true,
  linkcolor=blue!55!black,
  citecolor=blue!55!black,
  urlcolor=blue!55!black,
  breaklinks=true,
]{hyperref}

\def\BibTeX{{\rm B\kern-.05em{\sc i\kern-.025em b}\kern-.08em
    T\kern-.1667em\lower.7ex\hbox{E}\kern-.125emX}}

\begin{document}

\title{Improving Text-to-Music Generation\\with Human Preference Rewards%
\thanks{Code \& Demo: \url{https://github.com/yonghyunk1m/ttm-humanpref}.}}

\author{%
\IEEEauthorblockN{%
  Yonghyun Kim\textsuperscript{1},~%
  Junwon Lee\textsuperscript{2},~%
  Haiwen Xia\textsuperscript{3},~%
  Yinghao Ma\textsuperscript{4},~%
  Chris Donahue\textsuperscript{5}%
}
\IEEEauthorblockA{%
  \textsuperscript{1}Georgia Tech~\quad
  \textsuperscript{2}KAIST~\quad
  \textsuperscript{3}Peking University~\quad
  \textsuperscript{4}Queen Mary University of London~\quad
  \textsuperscript{5}Carnegie Mellon University
}
}

\maketitle

\begin{abstract}
We describe our entry to the efficiency track of the Academic Text-to-Music (ATTM) Grand Challenge at ICME~$2026$. Beyond the challenge protocol's FAD-CLAP and CLAP score, we add a learned human-preference reward from \emph{TuneJury}, a twin pairwise ranker trained over open music-preference datasets. The reward serves both as a training-time conditioning signal and as a sample-selection criterion. The pipeline combines five engineering decisions on a $120$\,M-parameter FluxAudio-S backbone, four at training time and one at inference:
\begin{enumerate*}[label=(\roman*),itemjoin={{, }},itemjoin*={{, and }}]
    \item training-time reward conditioning that doubles as an inference-time CFG axis
    \item a sweep over five score-conditioning architectures, where training and inference use different variants
    \item expert iteration on the top decile
    \item a short preference-tuning pass (CRPO) for audio-text alignment
    \item inference post-processing via joint CFG, source separation, and loudness normalization.
\end{enumerate*}
Per-stage decomposition on $100$ Song Describer prompts shows training-time reward conditioning as a functional conditioning axis, expert iteration as the dominant contributor, the preference-tuning pass adding only noise-level gain, and the inference-time score scalar already saturated by the end of the chain.
\end{abstract}


\section{Introduction}
\label{sec:intro}

This paper reports our submission to the efficiency track ($\leq 500$\,M parameters) of the Academic Text-to-Music (ATTM) Grand Challenge at ICME~$2026$~\cite{hsieh2026academic}. The challenge protocol evaluates three objective metrics: FAD-CLAP (Fr\'echet Audio Distance~\cite{kilgour2019fad} on LAION-CLAP-Music audio embeddings~\cite{wu2023clap}) against the SDD-$706$ reference, a $706$-track instrumental subset of MTG-Jamendo~\cite{bogdanov2019mtg} from the Song Describer Dataset (SDD)~\cite{manco2023sdd}; CLAP score, the cosine similarity between the CLAP-text and CLAP-audio embeddings of each prompt-clip pair~\cite{wu2023clap}; and a Concept Coverage Score (CCS)~\cite{hsieh2026academic} computed by a large audio-language model judge. We focus on the first two in our internal tables and report the official CCS result in the Section~\ref{sec:results} footnote.

Beyond these two metrics, we use a learned human-preference reward supplied by \emph{TuneJury}~\cite{tunejury}, a twin pairwise ranker~\cite{burges2005ranknet} over LAION-CLAP-Music~\cite{wu2023clap} and MERT~\cite{li2024mert} features, trained on open music-preference datasets. The reward enters the pipeline in two roles: a per-clip training-time conditioning signal, and a selection criterion for self-generated samples used in the expert-iteration fine-tune.

Our submission combines five concrete engineering decisions on the $120$\,M-parameter FluxAudio-S baseline~\cite{fei2024fluxmusic,hsieh2026academic} provided by the challenge: four act on the backbone weights at training time, and one operates only at inference.

\smallskip\noindent\textit{Training-time decisions.}

\textbf{(i) Conditioning on the human-preference reward.} The per-clip TuneJury score enters the backbone as a Fourier-embedded~\cite{tancik2020fourier} side input. Score-conditioned variants improve FAD-CLAP by $0.025$--$0.040$ absolute over the unconditional baseline (Table~\ref{tab:method_comparison}). Null-score dropout makes the reward an additional classifier-free guidance (CFG)~\cite{ho2022cfg} axis at inference.

\textbf{(ii) Sweep over score-conditioning heads.} Of five injection heads on Jamendo-$100$, our $100$-clip MTG-Jamendo holdout (Table~\ref{tab:method_comparison}), InputAdd (v$2$) leads on FAD-CLAP, CLAP score, and input-score correlation. We deploy a v$1$ $\to$ v$2$ \emph{hybrid}: train Stages~$1$--$2$ in the more-stable GlobalAdaLN (v$1$) forward, then cross-load into the InputAdd (v$2$) forward at Stage~$3$. The reverse cross is unsafe (Section~\ref{sec:cross_ablation}).

\textbf{(iii) Reward-guided expert iteration~\cite{anthony2017thinking,gulcehre2023rest}.} We rank samples from the score-conditioned supervised fine-tuning (SFT) checkpoint by an equal-weight blend of ranker reward and CLAP-text similarity, and fine-tune on the top decile. This step is the dominant chain contributor: $-0.0362$ FAD-CLAP on the v$1$ chain (Row~1 $\to$ Row~2 of Table~\ref{tab:cumulative}).

\textbf{(iv) Short preference tuning for CLAP-text alignment.} A CLAP-Ranked Preference Optimization (CRPO)~\cite{tangoflux2026} pass with a direct preference optimization (DPO)~\cite{rafailov2023dpo}-style objective fine-tunes the expert-iteration checkpoint on ${\sim}2$K CLAP-aligned winner/loser pairs. The delta over expert iteration alone is within paired-$t$ noise, but compute is negligible.

\smallskip\noindent\textit{Inference-time decision.}

\textbf{(v) Inference setup.} Joint CFG~\cite{ho2022cfg} on text and reward, a fixed prompt prefix, $3{\times}$Demucs~\cite{defossez2019demucs} \texttt{mdx\_extra} source separation, and LUFS normalization (Section~\ref{sec:inference}).

\medskip

The remainder of the paper expands on these decisions (Section~\ref{sec:method}) and reports per-stage ablations (Section~\ref{sec:results}).

\paragraph{Scope}
The present report is scoped to the engineering pipeline of the submission: at inference, we use a fixed single-value score scalar selected on SDD-$100$, a $100$-prompt subset we sampled from SDD for internal validation (Section~\ref{sec:results}) and do not analyze the score-response curve. Three analytical questions are left to future work: (a)~why reward-conditioned flow matching admits inference-time CFG \emph{extrapolation} past the reward scalar's training support, (b)~where this extrapolation breaks down, and (c)~how it generalizes to other backbones.

\paragraph{Workflow note}
The engineering reported here was carried out in a human-agent loop using Claude Code (Anthropic's Claude Opus $4.6$/$4.7$), in the spirit of AI-Driven Research for Systems~\cite{cheng2025barbarians}: the authors directed the design and ran every training/evaluation job, while the agent drafted and iterated on the implementation (code, scripts, and manuscript) under the authors' review. The full human-agent split is detailed in \nameref{sec:ai_disclosure}.

\section{Related Work}
\label{sec:related}

Our pipeline draws on a small set of well-established ingredients from the flow-matching, preference-learning, and music-generation literature. We close with a brief note on AI-driven research workflows, the methodological frame for our engineering loop.

\paragraph{Flow matching for audio generation}
\emph{Flow matching}~\cite{lipman2023flow} learns a continuous-time velocity field whose Euler integration maps prior noise to data. The Flux-style flow-matching transformer~\cite{fei2024fluxmusic} provides our backbone, in the form of the FluxAudio-S baseline supplied by the challenge organizers~\cite{hsieh2026academic}. \emph{Classifier-free guidance}~\cite{ho2022cfg} combines a conditional and an unconditional pass at inference to amplify the conditioning signal.

\paragraph{Self-improvement via expert iteration}
\emph{Expert iteration}, in the ExIt~\cite{anthony2017thinking} and ReST~\cite{gulcehre2023rest} formulations, alternates between sampling from the current policy and fine-tuning on top-quality samples. We use a one-round version filtered by our learned reward jointly with CLAP-text similarity.

\paragraph{Preference optimization}
\emph{Direct preference optimization} (DPO)~\cite{rafailov2023dpo} fits a policy directly to pairwise preferences without training a separate reward model. \emph{CLAP-Ranked Preference Optimization} (CRPO), introduced in TangoFlux~\cite{tangoflux2026}, adapts DPO to text-to-music by constructing preference pairs with a CLAP-text scorer, and we use the same procedure.

\paragraph{Music representations and preference data}
Music-audio encoders include text-audio contrastive models (e.g., LAION-CLAP-Music~\cite{wu2023clap}) and music-pretrained self-supervised models (e.g., MERT-v$1$-$330$M~\cite{li2024mert}). Open preference data for music has emerged across multiple sources, including Music Arena~\cite{kim2025musicarena}, MusicPrefs~\cite{huang2025musicprefs}, AIME~\cite{grotschla2025aime}, and SongEval~\cite{yao2025songeval}. Our ranker pools all four with a RankNet~\cite{burges2005ranknet} pairwise logistic loss (Section~\ref{sec:rw}).

\paragraph{AI-driven research workflows}
The pattern of a human-defined objective with an LLM agent iterating against a programmatic evaluator has recently been formalized as AI-Driven Research for Systems~\cite{cheng2025barbarians}, with closely related instances in FunSearch~\cite{romeraparedes2024funsearch} (math) and AlphaEvolve~\cite{novikov2025alphaevolve} (algorithms). Our human-agent loop borrows the same high-level structure, with full disclosure in \nameref{sec:ai_disclosure}.

\section{Proposed Method}
\label{sec:method}

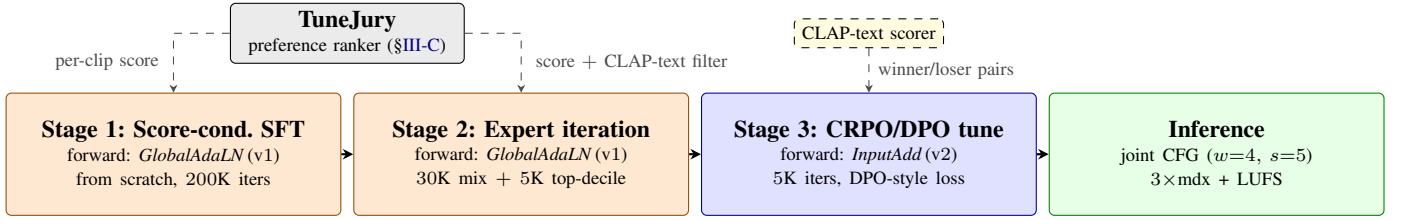
\begin{figure*}[!t]
\centering
\begin{tikzpicture}[
  font=\small,
  every node/.style={align=center},
  block/.style={
    rectangle, draw, rounded corners=2pt,
    minimum height=46pt, text width=118pt, align=center,
    inner sep=4pt
  },
  ranker/.style={
    rectangle, draw, rounded corners=2pt, fill=black!8,
    align=center, inner sep=3pt, font=\small,
    minimum height=22pt, minimum width=88pt
  },
  stagev1/.style={block, fill=orange!18, draw=orange!60!black},
  stagev2/.style={block, fill=blue!12,   draw=blue!60!black},
  infer/.style ={block, fill=green!10,  draw=green!50!black},
  oracle/.style={
    rectangle, draw, dashed, rounded corners=2pt,
    fill=yellow!15, inner sep=3pt, font=\scriptsize, align=center
  },
  flow/.style={-{Stealth[length=4pt,width=4pt]}, thick},
  signal/.style={-{Stealth[length=3pt,width=3pt]}, thin, dashed, gray!60!black},
  lbl/.style={font=\scriptsize, align=center},
]

% --- Bottom row: 4 stages, uniform width (text width=118pt), gap 4.6 ---
\node[stagev1] (s1) at (0,0)    {%
  \textbf{Stage 1: Score-cond.\ SFT}\\[1pt]
  \scriptsize forward: \emph{GlobalAdaLN}\,(v$1$)\\[-1pt]
  \scriptsize from scratch, $200$K iters};

\node[stagev1] (s2) at (4.6,0)  {%
  \textbf{Stage 2: Expert iteration}\\[1pt]
  \scriptsize forward: \emph{GlobalAdaLN}\,(v$1$)\\[-1pt]
  \scriptsize $30$K mix $+$ $5$K top-decile};

\node[stagev2] (s3) at (9.2,0)  {%
  \textbf{Stage 3: CRPO/DPO tune}\\[1pt]
  \scriptsize forward: \emph{InputAdd}\,(v$2$)\\[-1pt]
  \scriptsize $5$K iters, DPO-style loss};

\node[infer]  (inf) at (13.8,0) {%
  \textbf{Inference}\\[1pt]
  \scriptsize joint CFG ($w{=}4,\,s{=}5$)\\[-1pt]
  \scriptsize $3{\times}$mdx + LUFS};

% --- Top: ranker and clap-text scorer, lowered closer to stages ---
\node[ranker] (ranker) at (2.3, 1.6) {%
  \textbf{TuneJury}\\[-1pt]
  \scriptsize preference ranker (\S\ref{sec:rw})};

\node[oracle] (clap) at (9.2, 1.6) {CLAP-text scorer};

% --- Stage chain (horizontal solid arrows; labels omitted to avoid
% overlap with the narrow inter-box gaps. The v$1\to$v$2$ transition
% is conveyed by the orange/blue/green fill colors and the caption.) ---
\draw[flow] (s1) -- (s2);
\draw[flow] (s2) -- (s3);
\draw[flow] (s3) -- (inf);

% --- Ranker -> Stage 1: leaves the LEFT side of the ranker, drops to s1 ---
\draw[signal] (ranker.west) -| (s1.north)
  node[pos=0.74, left=2pt, lbl, gray!60!black]{per-clip score};

% --- Ranker -> Stage 2: leaves the RIGHT side of the ranker, drops to s2 ---
\draw[signal] (ranker.east) -| (s2.north)
  node[pos=0.74, right=2pt, lbl, gray!60!black]{score $+$ CLAP-text filter};

% --- CLAP-text -> Stage 3 (vertical) ---
\draw[signal] (clap) -- (s3.north)
  node[midway, right, lbl, gray!60!black]{winner/loser pairs};

\end{tikzpicture}
\caption{\textbf{End-to-end system pipeline.} Box color marks the
score-conditioning forward in use: orange for \emph{GlobalAdaLN
(v$1$)} (Stages~$1$ and~$2$), blue for \emph{InputAdd (v$2$)}
(Stage~$3$), and green for the deployed Inference endpoint
(inherits v$2$ from Stage~$3$). GlobalAdaLN modulates the AdaLN
parameters of every transformer block, and InputAdd broadcasts
the reward embedding to every audio latent at the input projection
only. Stages~$1$ and~$2$ train in the v$1$ forward because v$1$
converged more stably in pilots, and Stage~$3$ cross-loads the v$1$
weights into the v$2$ forward (parameter graphs are identical) and
runs CRPO/DPO. The TuneJury score (gray dashed) is the training-time
conditioning signal at Stage~$1$ and contributes to the top-decile
filter at Stage~$2$, while the CRPO winner/loser pairs at Stage~$3$
are constructed from CLAP-text alignment (yellow dashed), following
the standard CRPO procedure.}
\label{fig:pipeline}
\end{figure*}

We describe the backbone, score-conditioning head, and preference ranker in this section, the training pipeline in Section~\ref{sec:training}, and the inference procedure in Section~\ref{sec:inference}. The deployed system trains as v$1$ (Stages~$1$--$2$) and switches to v$2$ only at Stage~$3$ via cross-loading, justified in Section~\ref{sec:cross_ablation}. Fig.~\ref{fig:pipeline} summarizes the pipeline.

% Pipeline overview figure (TikZ).
% Usage in main paper: \input{figures/pipeline}
%
% Required packages (declare in main preamble if not already):
%   \usepackage{tikz}
%   \usetikzlibrary{arrows.meta, positioning, shapes.geometric, fit, calc}

\subsection{Backbone}
\label{sec:backbone}

The generative backbone is \emph{FluxAudio-S}, the $120$\,M-parameter Flux-style flow-matching transformer~\cite{fei2024fluxmusic}, with the unconditional checkpoint released by MeanAudio~\cite{li2025meanaudio} designated by the challenge as the efficiency-track baseline~\cite{hsieh2026academic}. It operates on $1$D-Mel variational autoencoder (VAE) latents at $44.1$\,kHz (${\sim}10$\,s per clip), with text conditioning via T5-Large~\cite{raffel2020t5} cross-attention and pooled LAION-CLAP~\cite{wu2023clap} features through adaptive layer normalization (AdaLN). Audio is synthesized from latents by a pretrained BigVGAN vocoder~\cite{lee2023bigvgan}. We adopt the FluxAudio-S architecture, add the score-conditioning head described in Section~\ref{sec:cond} (no other layer modified), and train all backbone weights from scratch (Section~\ref{sec:training}). The MeanAudio-released checkpoint serves only as the Row~$0$ reference baseline in Table~\ref{tab:cumulative}.

\subsection{Score-Conditioning Head}
\label{sec:cond}

The reward scalar $s\in\mathbb{R}$ enters as a second conditioning input parallel to text. It is mapped to a $448$-d embedding $e_s$ via Fourier features~\cite{tancik2020fourier} and an MLP with a zero-initialized final projection, so the generator at the start of training is identical to the unconditional backbone. We compared five injection strategies on Jamendo-$100$, our $100$-clip MTG-Jamendo holdout (Table~\ref{tab:method_comparison}). InputAdd (v$2$), which broadcasts $e_s$ to every audio latent at the input projection ($z_i \leftarrow z_i + e_s$), leads on FAD-CLAP, CLAP score, and input-score correlation. The deployed model uses InputAdd (v$2$) at inference with weights warm-started from a GlobalAdaLN (v$1$) chain. v$1$ and v$2$ share an identical parameter graph and differ only in the forward roles of the score-related weights (Section~\ref{sec:cross_ablation}). The score is null-dropped ($\varnothing_s{=}0$) with probability $0.1$ during training, mirroring text CFG.

\begin{table}[t]
\caption{Score-conditioning architecture comparison on the
Jamendo-$100$ validation set (one generation per clip per input score). CLAP: CLAP-text cosine similarity. Score-$r$: Pearson correlation between input $s$ and output reward. $\Delta_{\text{out}}$: mean output reward at $s{=}{+}1.5$ minus that at $s{=}{-}0.5$ (higher = more steerable). Best per column in bold.}
\label{tab:method_comparison}
\centering
\setlength{\tabcolsep}{4pt}
\resizebox{\columnwidth}{!}{%
\begin{tabular}{lcccc}
\toprule
Variant & FAD-CLAP $\downarrow$ & CLAP $\uparrow$ & Score-$r$ $\uparrow$ & $\Delta_{\text{out}}$ \\
\midrule
FluxAudio-S (baseline) & $0.377$ & $0.213$ & --      & --      \\
GlobalAdaLN (v$1$)    & $0.352$ & $0.242$ & $0.442$ & $\mathbf{0.942}$ \\
InputAdd (v$2$)       & $\mathbf{0.337}$ & $\mathbf{0.249}$ & $\mathbf{0.524}$ & $0.779$ \\
AudioPrepend (v$3$)   & $0.339$ & $0.245$ & $0.439$ & $0.825$ \\
PerBlock AdaLN (v$4$) & $0.347$ & $0.243$ & $0.446$ & $0.856$ \\
TextPrepend (v$5$)    & $0.348$ & $0.244$ & $0.439$ & $0.757$ \\
\bottomrule
\end{tabular}}
\end{table}

\subsection{Pairwise Preference Ranker}
\label{sec:rw}

Our preference ranker, TuneJury~\cite{tunejury}, is a twin pairwise model that maps any audio clip plus an optional text prompt to a single quality scalar. In the released CLAP$+$MERT instantiation, each branch takes a $2048$-d concatenation of LAION-CLAP-Music~\cite{wu2023clap} audio ($512$), MERT-v$1$-$330$M~\cite{li2024mert} audio ($1024$), and LAION-CLAP-Music text ($512$). LAION-CLAP-Music supplies a caption-aligned semantic representation while MERT covers pitch, harmony, rhythm, and timbre that text-aligned encoders under-represent. The pairwise (rather than pointwise-regression) formulation matches the supervision: each of our four sources releases human votes as A-vs-B preferences, so the standard RankNet~\cite{burges2005ranknet} pairwise logistic loss $\mathcal{L} = -\log\sigma(s(A){-}s(B))$ consumes the labels directly. The score head is an MLP $2048{\to}1024{\to}512{\to}256{\to}128{\to}1$ with BatchNorm$+$ReLU$+$Dropout($0.5$), trained on ${\sim}22$K pairs (${\sim}2$K held out) pooled from Music Arena~\cite{kim2025musicarena}, MusicPrefs~\cite{huang2025musicprefs}, AIME~\cite{grotschla2025aime}, and SongEval~\cite{yao2025songeval}. Held-out pairwise accuracy is $70.3\%$, with expected calibration error (ECE)~$0.027$. We use the ranker in two roles within the pipeline: (a)~as a per-clip quality score that we attach to every training example and feed to the score-conditioning head, and (b)~as the filter (jointly with a CLAP-text similarity score) that selects the top decile of self-generated samples for the expert-iteration fine-tune (Section~\ref{sec:training}). The CRPO preference-tuning pass constructs its winner/loser pairs by CLAP-text alignment alone, following the standard CRPO procedure. Full design-space ablations are in the released repository~\cite{tunejury}.

\subsection{Training Pipeline}
\label{sec:training}

The four training-time decisions of Section~\ref{sec:intro} are implemented as a three-stage chain: Stage~$1$ trains the score-conditioned backbone (operationalizing decisions~(i) and (ii)), Stage~$2$ runs (iii)~expert iteration, and Stage~$3$ runs (iv)~CRPO. All three stages train on the same Demucs-separated instrumental stem of MTG-Jamendo, and only the data weighting and the loss change between stages.

\paragraph{Data}

We start from the challenge-provided \texttt{jamendo\_qwen.json} captions over the ${\sim}55$K-track MTG-Jamendo dataset~\cite{bogdanov2019mtg}, segment audio into $10$\,s clips (${\sim}535$K clips), and apply Demucs vocal separation to keep the instrumental stem only. Three reward columns per clip are computed with the ranker: \texttt{reward\_score} (clip-level on full audio), \texttt{instrumental\_reward\_score} (clip-level on instrumental audio), and \texttt{track\_reward\_score} (track-level mean). The submitted model uses \texttt{instrumental\_reward\_score} as the conditioning signal because mixed-audio scoring partly tracks vocal presence. The generator learns to insert vocal-like artifacts to inflate the reward, hurting FAD-CLAP against an instrumental reference (FAD-CLAP $0.515$ vs. $0.337$ for two SFT runs trained with full-mix reward and instrumental-stem reward, respectively, under otherwise identical hyperparameters). Train-split statistics for the instrumental reward score are mean~$0.62$, std~$0.59$, $p_5{=}{-}0.45$, $p_{95}{=}{+}1.46$, $\max{=}{+}2.76$. Validation and test splits hold out $200$ tracks each; the Jamendo-$100$ ablation set (Section~\ref{sec:cond}) is a $100$-clip subset of this validation split. The original MTG-Jamendo tag vocabulary is not used directly: tag-derived genre/instrument/mood words appear inside these captions and reach the model only through the natural-language path.

\paragraph{Stage 1 (Score-Conditioned SFT)}

Stage~1 trains a score-conditioned backbone from scratch on the full ${\sim}535$K-clip set in the GlobalAdaLN (v$1$) forward, the most stable variant under our budget despite InputAdd (v$2$)'s slight edge in Table~\ref{tab:method_comparison}. v$1$ carries through Stage~2 and is cross-loaded into v$2$ at Stage~3 (Section~\ref{sec:cross_ablation}). Hyperparameters: AdamW, lr $10^{-4}$ (constant after a $1$K-step warmup), effective batch $64$, bf$16$, score-null dropout $0.1$, $200$K updates (${\sim}32$\,h on one NVIDIA RTX~A$5000$), with EMA at $\sigma_{\text{rel}}\!\in\!\{0.05, 0.1\}$.

\paragraph{Stage 2 (Expert Iteration)}

Stage~2 fine-tunes the SFT checkpoint on a top-decile filter of its own outputs, in the spirit of expert iteration~\cite{anthony2017thinking,gulcehre2023rest}. We sample ${\sim}630$ clips from the SFT checkpoint at $s{=}2.0$, rank them by an equal-weight $z$-score blend of ranker reward and CLAP-text similarity, and keep the top decile ($64$ clips, reward mean $+1.05$, comparable to the upper ${\sim}20\%$ of the training distribution). The kept clips are then $5{\times}$-oversampled into the ${\sim}535$K-clip mixture and the checkpoint is fine-tuned for $30$K steps at lr $10^{-5}$, followed by a brief $5$K-step polish on the top-decile subset at lr $10^{-6}$.

\paragraph{Stage 3 (CRPO Preference-Tuning)}

Stage~3 switches to InputAdd (v$2$) and warm-starts backbone$+$\texttt{score\_embed} from the v$1$ expert-iteration checkpoint via shape-matched partial loading that transfers all $203$ keys (Section~\ref{sec:cross_ablation}). It then runs CRPO~\cite{tangoflux2026} on $2{,}000$ preference pairs: we score generated samples by CLAP-text alignment under each prompt and pair each high-CLAP sample with a low-CLAP sample under the same prompt. The DPO~\cite{rafailov2023dpo}-style loss is
\begin{equation}
\mathcal{L}_{\text{CRPO}}
 = -\log\sigma\!\Bigl(\beta\bigl(\Delta_{\text{win}}^{\pi}
                                - \Delta_{\text{lose}}^{\pi}\bigr)\Bigr)
   + \lambda_{\text{FM}}\,\mathcal{L}_{\text{FM}}^{\text{win}}
\label{eq:crpo}
\end{equation}
with $\Delta_{x}^{\pi} = \log[\pi(x)/\pi_{\text{ref}}(x)]$, $\beta{=}2000$ (the large $\beta$ matches TangoFlux's CRPO scaling for flow-matching log-likelihood ratios, which are larger in magnitude than language-model token log-probs), $\lambda_{\text{FM}}{=}1.0$, lr $10^{-6}$, $5$K updates. The flow-matching auxiliary $\mathcal{L}_{\text{FM}}^{\text{win}}$ regularizes toward the warm-started reference. The resulting checkpoint is the submitted model.

\paragraph{Total compute} The full pipeline (SFT $+$ expert-iteration
$+$ CRPO $+$ ranker training) fits in approximately $40$ GPU-hours on a single NVIDIA RTX A$5000$.

\subsection{Inference Procedure}
\label{sec:inference}

\paragraph{Joint classifier-free guidance}
At inference, we apply classifier-free guidance~\cite{ho2022cfg} jointly on text and reward:
\begin{equation}
\tilde v = v(x_t, t, \varnothing_t, \varnothing_s)
 + w\bigl[ v(x_t, t, c, s) - v(x_t, t, \varnothing_t, \varnothing_s)\bigr],
\label{eq:joint_cfg}
\end{equation}
where $\varnothing_t$ is the text-null, $\varnothing_s{=}0$ the score-null, and the same scalar $w$ lifts text and score conditioning jointly relative to the doubly-unconditional baseline. We hold $w$ fixed at $4.0$ and the score scalar fixed at $s{=}5.0$, both selected on the SDD-$100$ validation set. We use a single fixed value at inference, not a sweep, consistent with the scope stated in Section~\ref{sec:intro}. The chosen $s{=}5.0$ lies above the training-time range of the reward score (max $+2.76$ on the train split, Section~\ref{sec:training}). A full analytical study of the score-response curve under extrapolation is left to future work. Sampling uses $25$ Euler steps (linear schedule $\sigma_i = 1 - i/25$), a seed fixed per submission, the prompt prefix ``\texttt{high quality instrumental music, }'', and a negative prompt (``\texttt{noise, distortion, low quality, static, hum, hiss, clipping, muffled, amateur recording}'') supplying $\varnothing_t$ in Eq.~\eqref{eq:joint_cfg} in place of an empty string. End-to-end wall-clock is $0.5$--$0.8$\,s per $10$\,s clip on a single NVIDIA RTX~A$5000$.

\paragraph{Source separation and loudness normalization}
Two lightweight post-processing steps consistently improve metrics on internal validation. First, we pass each generated wav through three sequential applications of Demucs's \texttt{mdx\_extra} model and keep the residual ``no-vocals'' track. Even with the ``\texttt{high quality instrumental music, }'' prefix, the score-conditioned generator occasionally produces vocal-like residuals that pollute FAD-CLAP against an instrumental reference, and the three-pass separator removes them. Second, we loudness-normalize the result to $-16.5$\,LUFS via the ITU-R BS.1770~\cite{itu2015bs1770} algorithm with a true-peak ceiling at $-1$\,dB. The LUFS target was selected on the validation set to minimize FAD-CLAP averaged across prompts, with comparable FAD-CLAP across the $-15$~to~$-18$~LUFS~range.

\paragraph{Submitted configurations}
We submit two configurations, Sub.~$1$ (seed $42$) and Sub.~$2$ (seed $55$), which share backbone weights and the post-processing pipeline above and differ only in the random seed used at inference.

\section{Experiments}
\label{sec:results}

We report internal validation on SDD-$100$, evaluated against the SDD-$706$ reference (both introduced in Section~\ref{sec:intro}). FAD-CLAP and CLAP score both use the LAION-CLAP-Music checkpoint \texttt{music\_audioset\_epoch\_15\_esc\_90.14.pt} on $10$-second clips, matching the official objective-metric protocol. FAD-CLAP is a distribution-level statistic (one value per condition); CLAP score and Reward are per-prompt and support paired-$t$ tests. Throughout this section, \emph{Reward} denotes the mean output of our preference ranker (Section~\ref{sec:rw}), and unless noted otherwise all rows use the same single-value inference protocol ($s{=}5.0$, $w{=}4.0$, $25$ Euler steps, prefix prompt, seed $42$ unless stated, $3{\times}$mdx\_extra, $-16.5$ LUFS). This SDD-$706$ protocol is distinct from the architecture-selection protocol of Table~\ref{tab:method_comparison} (Jamendo-$100$ reference, no post-processing), and absolute values across the two are not directly comparable.\footnote{Under the challenge's hidden Jamendo reference set, our submission (\texttt{e02}) scored FAD $0.498$, CLAP $0.270$, CCS $0.763$~\cite{hsieh2026academic}.}

\subsection{Cumulative Stage Ablation}
\label{sec:cumulative}

Row $0$ in Table~\ref{tab:cumulative} is the MeanAudio-released unconditional FluxAudio-S checkpoint~\cite{li2025meanaudio} (generated without score conditioning), included only as a reference baseline. We add each pipeline step in order from Row $1$ onward, measuring the marginal contribution against the previous row. Only step~$2$ (expert iteration) reaches paired-$t$ significance on either CLAP score or Reward, while steps~$3$ and~$4$ each leave the per-prompt distribution within paired-$t$ noise of the previous row, in agreement with the cross-mechanism findings.

\begin{table}[t]
\caption{Cumulative ablation along the deployed chain ($N{=}100$
SDD prompts). Sub.~$2$ is the seed-$55$ sibling of Sub.~$1$ (same chain). $^{\dagger}$: paired-$t$ improvement over the previous row (one-sided, $p{<}0.05$). Best per column in bold.}
\label{tab:cumulative}
\centering
\setlength{\tabcolsep}{4pt}
\begin{tabular}{clccc}
\toprule
\# & Pipeline (cumulative) & FAD-CLAP $\downarrow$ & CLAP $\uparrow$ & Reward $\uparrow$ \\
\midrule
$0$ & FluxAudio-S (baseline)  & $0.5998$ & $0.230$ & $-0.392$ \\
\midrule
$1$ & Score-conditioned SFT (v$1$)        & $0.4681$ & $0.262$ & $+0.028$ \\
$2$ & $+$ Expert iteration                & $0.4319$ & $0.290^{\dagger}$ & $+0.524^{\dagger}$ \\
$3$ & $+$ Cross-load to v$2$ forward      & $0.4272$ & $0.283$ & $+0.535$ \\
$4$ & $+$ CRPO ($=$ Sub.~$1$, seed $42$)  & $\mathbf{0.4238}$ & $0.285$ & $+0.533$ \\
   & \quad Sub.~$2$ (seed $55$)           & $0.4370$ & $\mathbf{0.300}$ & $\mathbf{+0.550}$ \\
\bottomrule
\end{tabular}
\end{table}

\subsection{Cross-Mechanism Ablation}
\label{sec:cross_ablation}

We treat the score-conditioning mechanism as a separable knob from the trained weights and run an $8$-cell factorial: \{\emph{SFT-only}, \emph{Chain-end}\} $\times$ \{v1 weights, v2 weights\} $\times$ \{v1 forward, v2 forward\}, plus the two submitted configurations (Table~\ref{tab:cross_ablation}). \emph{SFT-only} is the post-Stage-1 checkpoint, \emph{Chain-end} is post-Stage-2 (before CRPO), and \emph{Hybrid (submitted)} is post-Stage-3 (CRPO over Chain-end v1 $\to$ v2). Cross-loading uses \texttt{state\_dict.load(strict=False)}, and v1 and v2 share an identical $203$-key parameter graph.

\begin{table}[t]
\caption{Cross-mechanism ablation ($N{=}100$ SDD prompts).
$^{*}$ marks cells statistically tied with Sub.~$1$ on the per-prompt margin (paired-$t$, $p\geq 0.05$). Unmarked cells are significantly worse than Sub.~$1$ on that metric. Best per column in bold. Hybrid rows are Sub.~$1$ (seed $42$) and its seed-$55$ sibling Sub.~$2$.}
\label{tab:cross_ablation}
\centering
\setlength{\tabcolsep}{4pt}
\resizebox{\columnwidth}{!}{%
\begin{tabular}{llccc}
\toprule
Stage & Weights $\to$ Forward & FAD-CLAP $\downarrow$ & CLAP $\uparrow$ & Reward $\uparrow$ \\
\midrule
\multirow{4}{*}{SFT-only} & v1 $\to$ v1 (native)  & $0.4681$ & $0.262$ & $+0.028$ \\
                          & v1 $\to$ v2 (cross)   & $0.4456$ & $0.265$ & $+0.009$ \\
                          & v2 $\to$ v1 (cross)   & $0.6846$ & $0.202$ & $-0.500$ \\
                          & v2 $\to$ v2 (native)  & $0.4442$ & $0.266$ & $+0.282$ \\
\midrule
\multirow{4}{*}{Chain-end} & v1 $\to$ v1 (native) & $0.4319$ & $0.290^{*}$ & $+0.524^{*}$ \\
                            & v1 $\to$ v2 (cross)  & $0.4272$ & $0.283^{*}$ & $+0.535^{*}$ \\
                            & v2 $\to$ v1 (cross)  & $0.6952$ & $0.198$ & $-0.518$ \\
                            & v2 $\to$ v2 (CRPO) & $0.4695$ & $0.265$ & $+0.244$ \\
\midrule
\multirow{2}{*}{Hybrid (submitted)}
   & v1 $\to$ v2 (Sub.~1, seed $42$) & $\mathbf{0.4238}$ & $0.285$          & $+0.533$ \\
   & v1 $\to$ v2 (Sub.~2, seed $55$) & $0.4370$          & $\mathbf{0.300}$ & $\mathbf{+0.550}$ \\
\bottomrule
\end{tabular}}
\end{table}

\subsection{Inference-Time Score Sensitivity}
\label{sec:score_sweep}

To check whether the inference-time score scalar does visible work on the deployed Hybrid (Sub.~$1$) checkpoint, we sweep $s\in[0,6]$ on the SFT-only and Hybrid checkpoints.

% Inference-time score sweep (Fig. 2). Pure pgfplots / TikZ, so the
% figure inherits the paper's IEEEtran serif font.
\begin{figure}[t]
\centering
\begin{tikzpicture}
  \pgfplotsset{
    every axis/.style={
      width=0.95\columnwidth,
      height=4.0cm,
      tick pos=left,
      grid=major,
      grid style={gray!22, dashed, line width=0.3pt},
      tick label style={font=\scriptsize},
      label style={font=\footnotesize},
      legend style={
        font=\scriptsize,
        cells={anchor=west},
        draw=gray!50,
        fill=white,
        fill opacity=0.92,
        text opacity=1,
        inner sep=2pt,
      },
      no markers=false,
      mark size=1.7pt,
      line width=0.9pt,
    },
    sft style/.style ={color=orange!85!red, mark=*},
    sub style/.style ={color=blue!75!black,  mark=square*},
  }
  % NOTE: the s=5 vertical guide is drawn with an inline style below.
  % Defining it in \pgfplotsset{} above does not work because
  % \draw[...] is a TikZ command and looks up styles in the /tikz/
  % namespace, not /pgfplots/, so the color silently falls back to
  % default black.

  % --- FAD panel (top) -----------------------------------------------------
  \begin{axis}[
    name=fadax,
    ylabel={FAD-CLAP\,$\downarrow$},
    xtick={0,1,2,3,4,5,6},
    xticklabels={},
    ymin=0.39, ymax=0.53,
    legend pos=north west,
    legend style={font=\scriptsize, at={(0.02,0.98)}, anchor=north west},
  ]
    \draw[black!30, densely dashed, line width=0.4pt]
      (axis cs:5,0.39) -- (axis cs:5,0.53);
    \node[font=\scriptsize, black!55, anchor=south west]
      at (axis cs:5.05,0.505) {$s{=}5$};

    \addplot+[sft style] coordinates {
      (0.0,0.4055) (1.0,0.4038) (2.0,0.4224) (3.0,0.4109)
      (4.0,0.4585) (4.5,0.4460) (5.0,0.4764) (5.5,0.4927)
      (6.0,0.5113)
    };
    \addlegendentry{SFT-only}

    \addplot+[sub style] coordinates {
      (0.0,0.4269) (1.0,0.4339) (2.0,0.4299) (3.0,0.4384)
      (4.0,0.4300) (4.5,0.4314) (5.0,0.4219) (5.5,0.4321)
      (6.0,0.4303)
    };
    \addlegendentry{Submitted (Hybrid)}
  \end{axis}

  % --- Reward panel (bottom) ----------------------------------------------
  \begin{axis}[
    at={(fadax.below south west)}, anchor=north west, yshift=-2pt,
    xlabel={reward-score CFG scale\ \,$s$},
    ylabel={Reward\,$\uparrow$},
    xtick={0,1,2,3,4,5,6},
    ymin=0.10, ymax=0.60,
  ]
    \draw[black!30, densely dashed, line width=0.4pt]
      (axis cs:5,0.10) -- (axis cs:5,0.60);

    \addplot+[sft style] coordinates {
      (0.0,0.1633) (1.0,0.2360) (2.0,0.4704) (3.0,0.4250)
      (4.0,0.2830) (4.5,0.2713) (5.0,0.2362) (5.5,0.1986)
      (6.0,0.1201)
    };

    \addplot+[sub style] coordinates {
      (0.0,0.5413) (1.0,0.5052) (2.0,0.5017) (3.0,0.5421)
      (4.0,0.5332) (4.5,0.5257) (5.0,0.5278) (5.5,0.5114)
      (6.0,0.5276)
    };
  \end{axis}
\end{tikzpicture}
\caption{\textbf{Inference-time score sweep on $100$ SDD prompts.}
\textbf{SFT-only} (orange) tracks the reward monotonically within
its training range: Spearman $\rho{=}1.0$ on Reward across
$s\in[0,2]$, with Reward rising from $+0.16$ to $+0.47$, and
past $s{=}3$ the curve bends and FAD-CLAP rises. \textbf{Submitted}
(blue, Hybrid Sub.~$1$ from
Table~\ref{tab:cross_ablation}) is essentially flat in both
metrics across the full $s\in[0,6]$ range (Reward range $0.04$,
Pearson $r{\approx}0$), i.e.\ the inference knob has saturated.
Dotted line marks the deployed value $s{=}5$.}
\label{fig:score_sweep}
\end{figure}
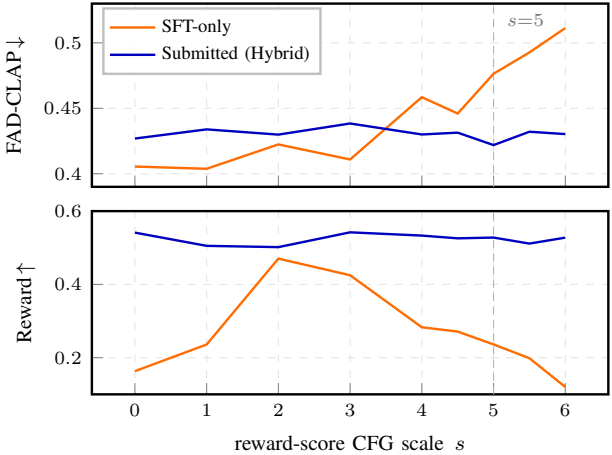

The two curves in Fig.~\ref{fig:score_sweep} diverge sharply. On the SFT-only checkpoint the score scalar moves Reward from $+0.16$ (at $s{=}0$) to $+0.47$ (at $s{=}2$) with Spearman $\rho{=}1.0$ across the training range $s\in[0,2]$, confirming that score conditioning at training time produced a backbone whose output reward tracks the input scalar monotonically. On the submitted Hybrid checkpoint, however, $s$ is nearly inert: Reward already sits at $+0.54$ at $s{=}0$, the entire $s\in[0,6]$ range varies Reward by less than $0.05$ and FAD-CLAP by less than $0.02$ (Pearson $r{\approx}0$ between $s$ and Reward), and the FAD-CLAP optimum at $s{=}5$ ($0.4219$) is within seed-noise of the unconditioned $s{=}0$ pass ($0.4269$). We hold $s{=}5.0$ in our submission because validation selected it, not because the inference-time score is the lever moving the model. The lever has been absorbed into the weights upstream by expert iteration and CRPO, which leaves the inference scalar with little remaining headroom.

\subsection{Engineering Observations}

We provide three takeaways from our development process.

\paragraph{Score conditioning matters at training time and is
saturated at inference} Score-conditioned variants improve FAD-CLAP by $0.025$--$0.040$ absolute over the unconditional baseline at the SFT stage (Table~\ref{tab:method_comparison}), but the deployed Hybrid checkpoint shows a flat $s$ response across $s\in[0,6]$ (Fig.~\ref{fig:score_sweep}, blue). The score signal therefore does its real work \emph{during} training: the backbone absorbs reward into its weights, and expert iteration and CRPO take up what little inference-time steerable margin remained. The chain in effect trades the SFT-only model's working $s$ knob for a higher absolute baseline, with Sub.~$1$ at any $s$ matching or exceeding the SFT-only peak.

\paragraph{Mechanism transfer is asymmetric: v1 $\to$ v2 is
benign, v2 $\to$ v1 collapses} Table~\ref{tab:cross_ablation} shows a sharp asymmetry. Loading a v1-trained checkpoint into the v2 (InputAdd) forward stays within $0.02$ Reward of the v1-native cell at both stages (Chain-end v$1$$\to$v$2$ Reward $+0.535$ vs.\ v$1$-native $+0.524$). The reverse cross collapses to FAD-CLAP~${\sim}0.69$ and Reward~${\sim}{-}0.50$. InputAdd is additive on audio tokens, so unfamiliar score weights dampen rather than distort. GlobalAdaLN modulates every layer, and a v2-trained \texttt{score\_embed} feeds the AdaLN with patterns far outside the training distribution. This justifies our hybrid direction of warm-starting a v2-architecture CRPO from a v1 backbone.

\paragraph{Expert iteration is the dominant chain contributor, while CRPO adds noise-level gain at this scale} The biggest delta in Table~\ref{tab:cross_ablation} comes from expert iteration on the v1 chain (SFT-only $\to$ Chain-end: FAD-CLAP $-0.0362$, CLAP $+0.028$, Reward $+0.496$). The v2 chain regresses on the same axis ($0.4442 \to 0.4695$, Reward $+0.282 \to +0.244$), as v2 expert iteration plus CRPO did not converge cleanly under our budget. Adding $5$K CRPO steps on top of the v$1$ chain (Chain-end v$1$ $\to$ v$2$ $\to$ Sub.~$1$) shifts FAD-CLAP by $-0.003$, CLAP by $+0.002$, and Reward by $-0.002$. Neither per-prompt difference reaches paired-$t$ significance at $p{<}0.05$.

\section{Conclusion}
\label{sec:conclusion}

We submitted a $40$~GPU-hour entry to the ICME~$2026$ ATTM Grand Challenge efficiency track on the $120$\,M-parameter FluxAudio-S baseline, with TuneJury supplying a learned human-preference reward used both as a training-time conditioning signal and as a sample-selection criterion. Three findings emerge from the per-stage ablation. (a)~Training-time reward conditioning is a functional steering axis (FAD-CLAP $0.025$--$0.040$ at SFT), but its effect is absorbed into the weights by chain-end and the inference-time scalar saturates. (b)~Mechanism transfer is asymmetric: v$1$ GlobalAdaLN $\to$ v$2$ InputAdd cross-loads benignly (and we deploy this hybrid), while the reverse collapses. (c)~Reward-filtered expert iteration is the dominant chain contributor ($-0.0362$ FAD-CLAP on the v$1$ chain), with the CRPO pass at noise-level gain. Future work includes an analytical study of the score-response curve under extrapolation and a cross-family replication that tests whether the findings transfer across architectures.

\section*{AI Workflow Disclosure}
\label{sec:ai_disclosure}

Building on the Workflow note in Section~\ref{sec:intro}, we record the human-agent split for transparency.

\paragraph{Tool} Claude Code CLI with Anthropic's
\emph{Claude Opus} $4.6$~\cite{anthropic_opus_4_6} and $4.7$~\cite{anthropic_opus_4_7}. Tasks were issued as conversational natural-language requests, and the agent had no autonomous evaluation budget against a fixed objective.

\paragraph{Direction (human)} All architectural, training-data,
evaluation, and post-processing choices reported above were proposed by the human authors, who also ran every training, generation, and evaluation job and validated the results. The agent's conceptual contribution was mostly mapping author-described procedures onto existing literature and suggesting baseline hyperparameters during review.

\paragraph{Implementation (agent)} The agent wrote most of the
line-level code (score-conditioning heads, expert-iteration sampling and filtering, CRPO loop, post-processing scripts, evaluation harnesses, figure source), drafted and edited the manuscript, maintained the bibliography, and tuned LaTeX layout.

\paragraph{Supervision} The authors reviewed every commit, revised manuscript edits, and gated all Overleaf pushes.

\bibliographystyle{IEEEbib}
\bibliography{icme2026references}

\end{document}